\documentstyle[12pt,epsfig]{article}                    %%%
\newcommand{\preprint}[1]{\begin{table}[t]      %%
            \begin{flushright}                  %%
            \begin{large}{#1}\end{large}        %%
            \end{flushright}                    %%
            \end{table}}                        %%
\begin{document}
\preprint{LA-UR-98-4357}
\title{Effect of Color Screening on Heavy Quarkonia Regge Trajectories}
\author{\\ M.M. Brisudov\'{a},\thanks{E-mail: BRISUDA@T5.LANL.GOV} \
L. Burakovsky\thanks{E-mail: BURAKOV@T5.LANL.GOV} \ and \ 
T. Goldman\thanks{E-mail: GOLDMAN@T5.LANL.GOV} \
\\  \\  Theoretical Division, MS B283 \\  Los Alamos National Laboratory \\ 
Los Alamos, NM 87545, USA }
%\date{ }
\maketitle
\begin{abstract}
Using an unquenched lattice potential to calculate the spectrum of a
bottomonium-like system, we demonstrate numerically that the effect of pair
creation is to produce termination of the real part of hadronic Regge 
trajectories, in 
contrast to the Veneziano model and the vast majority of phenomenological
generalizations. Termination of the real part of Regge trajectories may have 
significant
experimental consequences.
\end{abstract}
\bigskip
{\it Key words:} Regge trajectories, spectroscopy, potential models, color 
screening

PACS: 12.39.Pn, 12.40.Nn, 12.40.Yx, 12.90.+b
\bigskip
\bigskip
\bigskip

It is well known that  hadrons %composed of light $(u,d,s)$ quarks
populate near-linear Regge trajectories; i.e., the orbital momentum
$\ell $ of the state is proportional to the square of its mass: $\ell
=\alpha'M^2(\ell )+\alpha (0),$ where the slope $\alpha'$ depends weakly 
(for light $(u,d,s)$ quarks) on the flavor content of the states lying on 
the corresponding trajectory. This is based on the knowledge of the lowest 
lying states. What happens, however, for highly excited states 
remains model-dependent. In the Veneziano model for scattering 
amplitudes~\cite{venz} there are infinitely many excitations populating 
linear Regge trajectories.  The same picture of  Regge trajectories arises 
from a linear confining potential~\cite{kang}, and asymptotically for any 
potential  where the dominant long distance behavior is linear.

In this paper we argue that in QCD, due to the pair creation that screens 
the potential at large distances, Regge trajectories  become nonlinear 
and their real parts terminate. To illustrate this point, we consider a 
potential 
obtained 
in unquenched lattice QCD in ref.~\cite{lattice1}, and use it in the 
Schr\"{o}dinger equation to calculate spectra of heavy quarkonia, in 
particular, for a bottomonium-like system, for which the use of the 
potential, as well as a nonrelativistic calculation, is best justified.

One can argue that an exact treatment of the problem would consist in the 
analysis of a 
multichannel potential scattering model which is characterized by the 
simultaneous presence of, and communication between Hamiltonians for two types 
of channels: 
one is the Hamiltonian of ordinary two-particle scattering channels (in our  
case, the open two-meson decay channels (e.g., $B\bar{B}),$), and the other is 
the Hamiltonian of permanently confined channels (in our case, the bare 
$b\bar{b}$ system).
The scattering Hamiltonian  has an 
absolutely continuous spectrum on the positive real axis (and perhaps a 
finite number of negative energy bound states corresponding to meson-meson 
molecules), and the confining  
Hamiltonian  has 
only a point spectrum with an accumulation point at infinity. These two 
types of channels are connected in the full Hamiltonian for the multichannel 
system by off-diagonal local potentials.

 The problem was 
studied in general, nonrelativistically, by Dashen {\it et al.} \cite{Dash} for 
noninteracting scattering states. All of their results carry over unchanged for 
a class of scattering channel potentials \cite{Dash} which includes a 
Yukawa-type interaction. 
They found that the spectrum of the full 
Hamiltonian consists of 3 parts: a finite number of negative energy 
eigenvalues, a discrete set of positive energy eigenvalues, and an ``absolutely 
continuous spectrum'' on the remainder of the positive real axis. Some of the 
bound states may be embedded in the continuum. Interestingly 
enough, in the examples studied, the number of positive energy bound states of 
the full Hamiltonian 
is {\it finite}, notwithstanding the fact that the number of such states of 
the unperturbed confining Hamiltonian is infinite, and Dashen et al. expected 
this to hold in general.  Note that adding just one 
open channel is sufficient for this qualitative behaviour to develop.

 With this established, let us concentrate on the potential energy as it varies 
with separation for the lowest lying 
channel in the case of a coupling to a single open channel.

Proceeding out from zero separation, we expect to find that this energy 
corresponds 
to the value of the potential of the confining system at short distances, since 
the 
system cannot contain the additional (anti)quarks needed to produce the meson 
states without a significant energy cost. Conversely, as we proceed inward from 
infinite separation, we expect an attractive (except for particular quantum 
numbers) 
Yukawa potential between the meson-antimeson pair due to the exchange of a light 
meson between them in the t-channel. As we extend both considerations towards a 
matching point, we expect to find that the quark potential exceeds the 
meson one, due to the confining nature of the former. The quark potential 
dominates up to a separation at which the probability of light quark-antiquark 
pair creation becomes significant. Quantum 
effects smooth the match, but we may still expect the potential to be 
non-monotonic 
in the matching region -- overshooting the (two meson) threshold energy and then 
falling back below it again to reach the Yukawa value, as one moves out from the 
origin. A schematic picture of the 2-channel potential is shown in Figure 1.
Details will depend on the particular channel considered.

\begin{figure}[htb]
\centerline{\epsfig{file=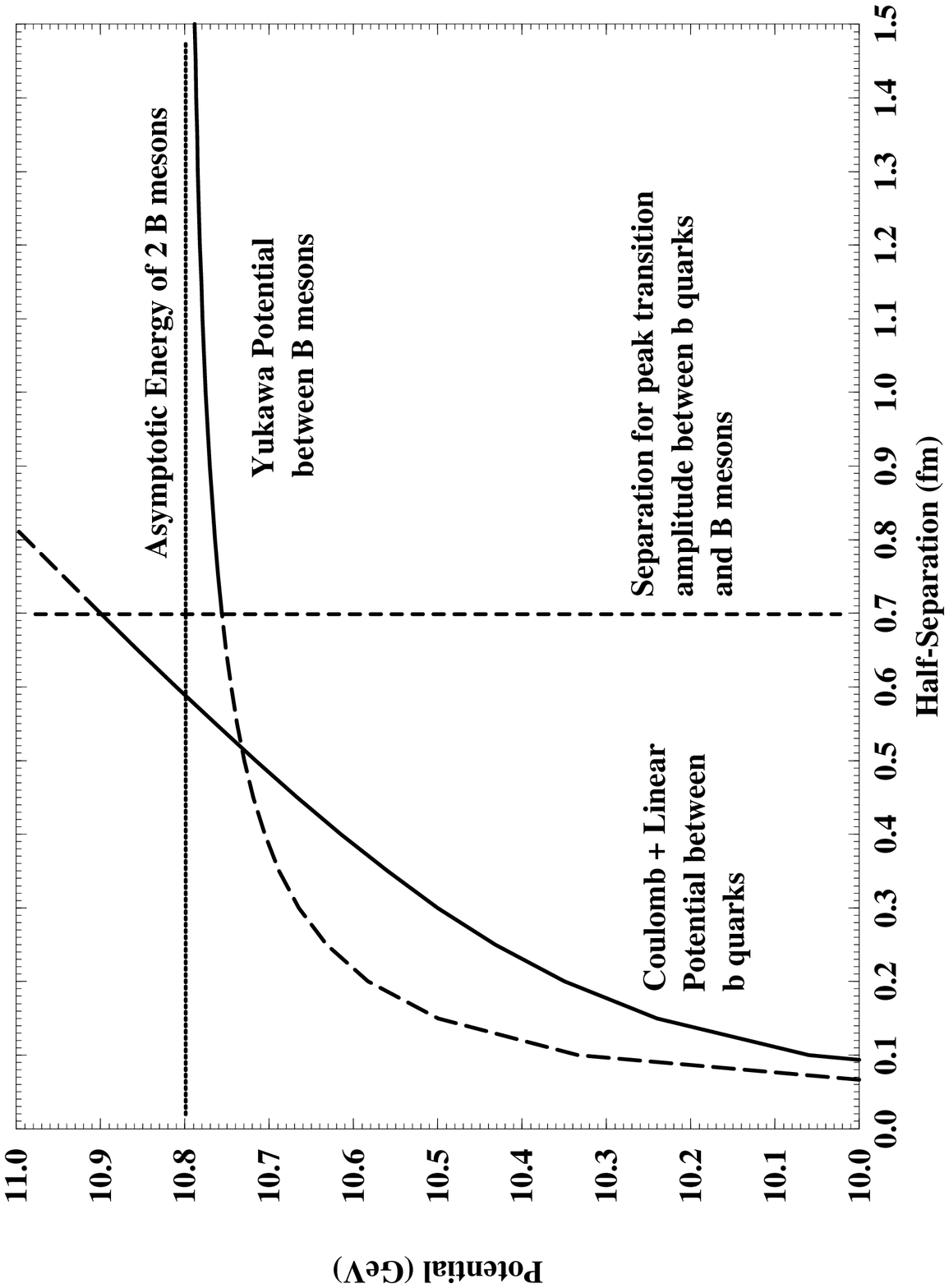,width=6in}} \vskip0.15in
%\caption{The non-relativistic quarkonium wave functions for the
%heavy ground state mesons $J/\psi$ and $\Upsilon$ from various potential
%models  and our calculation  (solid
%lines).  In the lower part of the figure, we also show a Gaussian fit
%adjusted to reproduce $\phi_V(k)$ at small $k$ (dotted lines). }
%fig_3}{6.0}
%\end{figure}
%\begin{figure}
%\input{terry2.ps}
\caption{{\bf A schematic representation of a two-coupled channel potential} 
using parameters similar to those in text. Yukawa and linear extensions are 
shown. }
\end{figure}
In fact, data from recent lattice 
calculations are consistent with this form of the effective potential, 
although it  may be that this
only appears to be so due to statistical fluctuations
\cite{potential}. Note that lattice studies include, in principle, a number of 
open channels.

We will show \'{a} posteriori  that using a two-body potential which follows the 
energy of the lowest channel (i.e. a static potential found in an unquenched 
lattice QCD) in a two-body calculation leads to qualitatively similar results to 
those of ref. \cite{Dash}, 
and argue that the two approaches should be expected to differ only near the 
threshold.  Note that the coupled-channel calculation would require more 
parameters than our two-body model where the coupling to open channels is 
included dynamically (through the lattice calculation).

Our purpose in this calculation is not to reproduce spectra; a better
agreement with data could be achieved by adjusting the quark mass
and/or fine tuning the parameters of the potential. Rather, we
concentrate on general features of the Regge trajectories that arise
for heavy quarkonia with a lattice QCD static potential once pair
production of light quarks is allowed.

For this purpose, it is sufficient for us to consider only the leading,
spin-independent part of the non-relativistic Hamiltonian, and argue
that since spin-dependent interactions are at least ${\cal O}(1/M^2)$
suppressed for heavy quarks, the general features of the trajectories
discussed below will persist in any more careful QCD bound-state
calculation. 

The bottomonium system is well described in leading order by a
nonrelativistic, spin-independent Hamiltonian, viz.
\begin{eqnarray}
H = -{1\over{2m}}{\boldmath {\nabla}^2 }+ V(r),
\end{eqnarray}
where $m=M/2$ is the reduced mass, with $M=5.2$ GeV, and  $V(r)$ denotes
a potential. Relativistic corrections in this case can be expected to
be on the order of 10\% ~\cite{lepage}. To make our calculation
represent QCD as closely as possible, we choose a potential that has been 
obtained in unquenched lattice QCD calculations for infinitely heavy
sources~\cite{lattice1}. One can expect that corrections to the
potential due to the  bottom mass being finite are small, of order
${\cal O}(\Lambda_{QCD}/M)$ (see~\cite{shuryak1} and references
therein).

The screened static potential fitted to results of lattice calculations
is~\cite{lattice1}:
\begin{eqnarray}
V(r) = \left( -{\alpha \over{r}} +\sigma r\right) {1 -e^{-\mu r} \over{\mu \, 
r}},
\end{eqnarray}
where $\mu ^{-1}=(0.9 \pm 0.2)$ fm $=(4.56 \pm 1.01)$ GeV$^{-1}$, 
$\sqrt{\sigma}=400$ MeV and $\alpha =0.21 \pm 0.01$. 
We obtain bottomonium spectra by diagonalizing the Hamiltonian $(1)$
with the potential $(2)$.

Before we proceed with the presentation of our results, a few comments regarding 
the lattice potential $(2)$ are called for.
First, the lattice screened potential in the analytic form $(2)$ does not
have the asymptotic Yukawa approach to the screened constant that
should occur as we argued above. Nevertheless, analytic results for
massless quarks~\cite{future}, which bracket Yukawa behavior, strongly
suggest that results qualitatively {\it and} quantitatively similar to
those presented below hold also in that more physical case.
Second, the potential $(2)$ is monotonic, unlike the qualitative picture shown 
schematically  in Fig. 1. The non-monotonic potential produces states which  mix 
strongly with scattering states, thus giving rise to resonances. This does not 
occur in the Hamiltonian with a monotonic potential (in the lowest order bound 
state pertubation theory). 
However, the drop of the potential value in the matched region can be expected 
to be at most tens of MeV (based on a typical strength of Yukawa potentials, and 
also consistent with the latest lattice results \cite{potential}). This means 
that the only states (not likely more than one or two) which could be 
significantly affected would be those that happen to lie in the tens-of-MeV-wide 
band around the 2 B threshold. We conclude that the potential $(2)$ is a 
sufficiently good approximation to the full coupled-channel problem for our 
purposes.

\begin{figure}[th!]
% GNUPLOT: LaTeX picture
\setlength{\unitlength}{0.240900pt}
\ifx\plotpoint\undefined\newsavebox{\plotpoint}\fi
\sbox{\plotpoint}{\rule[-0.200pt]{0.400pt}{0.400pt}}%
\begin{picture}(1500,1259)(0,0)
\font\gnuplot=cmr10 at 10pt
\gnuplot
\sbox{\plotpoint}{\rule[-0.200pt]{0.400pt}{0.400pt}}%
\put(220.0,164.0){\rule[-0.200pt]{292.934pt}{0.400pt}}
\put(220.0,164.0){\rule[-0.200pt]{4.818pt}{0.400pt}}
\put(198,164){\makebox(0,0)[r]{0}}
\put(1416.0,164.0){\rule[-0.200pt]{4.818pt}{0.400pt}}
\put(220.0,368.0){\rule[-0.200pt]{4.818pt}{0.400pt}}
\put(198,368){\makebox(0,0)[r]{2}}
\put(1416.0,368.0){\rule[-0.200pt]{4.818pt}{0.400pt}}
\put(220.0,572.0){\rule[-0.200pt]{4.818pt}{0.400pt}}
\put(198,572){\makebox(0,0)[r]{4}}
\put(1416.0,572.0){\rule[-0.200pt]{4.818pt}{0.400pt}}
\put(220.0,777.0){\rule[-0.200pt]{4.818pt}{0.400pt}}
\put(198,777){\makebox(0,0)[r]{6}}
\put(1416.0,777.0){\rule[-0.200pt]{4.818pt}{0.400pt}}
\put(220.0,981.0){\rule[-0.200pt]{4.818pt}{0.400pt}}
\put(198,981){\makebox(0,0)[r]{8}}
\put(1416.0,981.0){\rule[-0.200pt]{4.818pt}{0.400pt}}
\put(220.0,1185.0){\rule[-0.200pt]{4.818pt}{0.400pt}}
\put(198,1185){\makebox(0,0)[r]{10}}
\put(1416.0,1185.0){\rule[-0.200pt]{4.818pt}{0.400pt}}
\put(220.0,113.0){\rule[-0.200pt]{0.400pt}{4.818pt}}
\put(220,68){\makebox(0,0){114}}
\put(220.0,1216.0){\rule[-0.200pt]{0.400pt}{4.818pt}}
\put(372.0,113.0){\rule[-0.200pt]{0.400pt}{4.818pt}}
\put(372,68){\makebox(0,0){116}}
\put(372.0,1216.0){\rule[-0.200pt]{0.400pt}{4.818pt}}
\put(524.0,113.0){\rule[-0.200pt]{0.400pt}{4.818pt}}
\put(524,68){\makebox(0,0){118}}
\put(524.0,1216.0){\rule[-0.200pt]{0.400pt}{4.818pt}}
\put(676.0,113.0){\rule[-0.200pt]{0.400pt}{4.818pt}}
\put(676,68){\makebox(0,0){120}}
\put(676.0,1216.0){\rule[-0.200pt]{0.400pt}{4.818pt}}
\put(828.0,113.0){\rule[-0.200pt]{0.400pt}{4.818pt}}
\put(828,68){\makebox(0,0){122}}
\put(828.0,1216.0){\rule[-0.200pt]{0.400pt}{4.818pt}}
\put(980.0,113.0){\rule[-0.200pt]{0.400pt}{4.818pt}}
\put(980,68){\makebox(0,0){124}}
\put(980.0,1216.0){\rule[-0.200pt]{0.400pt}{4.818pt}}
\put(1132.0,113.0){\rule[-0.200pt]{0.400pt}{4.818pt}}
\put(1132,68){\makebox(0,0){126}}
\put(1132.0,1216.0){\rule[-0.200pt]{0.400pt}{4.818pt}}
\put(1284.0,113.0){\rule[-0.200pt]{0.400pt}{4.818pt}}
\put(1284,68){\makebox(0,0){128}}
\put(1284.0,1216.0){\rule[-0.200pt]{0.400pt}{4.818pt}}
\put(1436.0,113.0){\rule[-0.200pt]{0.400pt}{4.818pt}}
\put(1436,68){\makebox(0,0){130}}
\put(1436.0,1216.0){\rule[-0.200pt]{0.400pt}{4.818pt}}
\put(220.0,113.0){\rule[-0.200pt]{292.934pt}{0.400pt}}
\put(1436.0,113.0){\rule[-0.200pt]{0.400pt}{270.531pt}}
\put(220.0,1236.0){\rule[-0.200pt]{292.934pt}{0.400pt}}
\put(45,674){\makebox(0,0){$\ell $}}
\put(828,23){\makebox(0,0){$\boldmath M^2 [{\rm GeV}^2]$}}
\put(220.0,113.0){\rule[-0.200pt]{0.400pt}{270.531pt}}
\put(524,981){\makebox(0,0)[r]{$\mu=0.18$GeV}}
\put(568,981){\makebox(0,0){$+$}}
\put(354,164){\makebox(0,0){$+$}}
\put(563,266){\makebox(0,0){$+$}}
\put(724,368){\makebox(0,0){$+$}}
\put(872,470){\makebox(0,0){$+$}}
\put(988,572){\makebox(0,0){$+$}}
\put(1069,675){\makebox(0,0){$+$}}
\put(1145,777){\makebox(0,0){$+$}}
\put(1213,879){\makebox(0,0){$+$}}
\put(1247,981){\makebox(0,0){$+$}}
\put(1249,1083){\makebox(0,0){$+$}}
\put(1251,1185){\makebox(0,0){$+$}}
\sbox{\plotpoint}{\rule[-0.400pt]{0.800pt}{0.800pt}}%
\put(524,936){\makebox(0,0)[r]{$\mu=0.22$GeV}}
\put(568,936){\rule{1pt}{1pt}}
\put(338,164){\rule{1pt}{1pt}}
\put(527,266){\rule{1pt}{1pt}}
\put(673,368){\rule{1pt}{1pt}}
\put(796,470){\rule{1pt}{1pt}}
\put(877,572){\rule{1pt}{1pt}}
\put(942,675){\rule{1pt}{1pt}}
\put(979,777){\rule{1pt}{1pt}}
\put(985,879){\rule{1pt}{1pt}}
\put(990,981){\rule{1pt}{1pt}}
\put(996,1083){\rule{1pt}{1pt}}
\put(1003,1185){\rule{1pt}{1pt}}
\sbox{\plotpoint}{\rule[-0.500pt]{1.000pt}{1.000pt}}%
\put(524,891){\makebox(0,0)[r]{$\mu=0.28$GeV}}
\put(568,891){\rule{1pt}{1pt}}
\put(314,164){\rule{1pt}{1pt}}
\put(476,266){\rule{1pt}{1pt}}
\put(597,368){\rule{1pt}{1pt}}
\put(676,470){\rule{1pt}{1pt}}
\put(712,572){\rule{1pt}{1pt}}
\put(716,675){\rule{1pt}{1pt}}
\put(720,777){\rule{1pt}{1pt}}
\put(725,879){\rule{1pt}{1pt}}
\put(730,981){\rule{1pt}{1pt}}
\put(736,1083){\rule{1pt}{1pt}}
\put(743,1185){\rule{1pt}{1pt}}
\sbox{\plotpoint}{\rule[-0.200pt]{0.400pt}{0.400pt}}%
\put(524,846){\makebox(0,0)[r]{"threshold"}}
\put(546.0,846.0){\rule[-0.200pt]{15.899pt}{0.400pt}}
\put(1243,164){\usebox{\plotpoint}}
\put(1243.0,164.0){\rule[-0.200pt]{0.400pt}{258.245pt}}
\put(971,164){\usebox{\plotpoint}}
\put(971.0,164.0){\rule[-0.200pt]{0.400pt}{245.959pt}}
\put(704,164){\usebox{\plotpoint}}
\put(704.0,164.0){\rule[-0.200pt]{0.400pt}{123.100pt}}
\end{picture}
\caption{The parent trajectory for bottomonium with the lattice
unquenched potential $(2)$ for various allowed values of $\mu$. Solid
vertical lines represent the ionization level, $(2M_b +\sigma / \mu)^2$.
}
\end{figure}
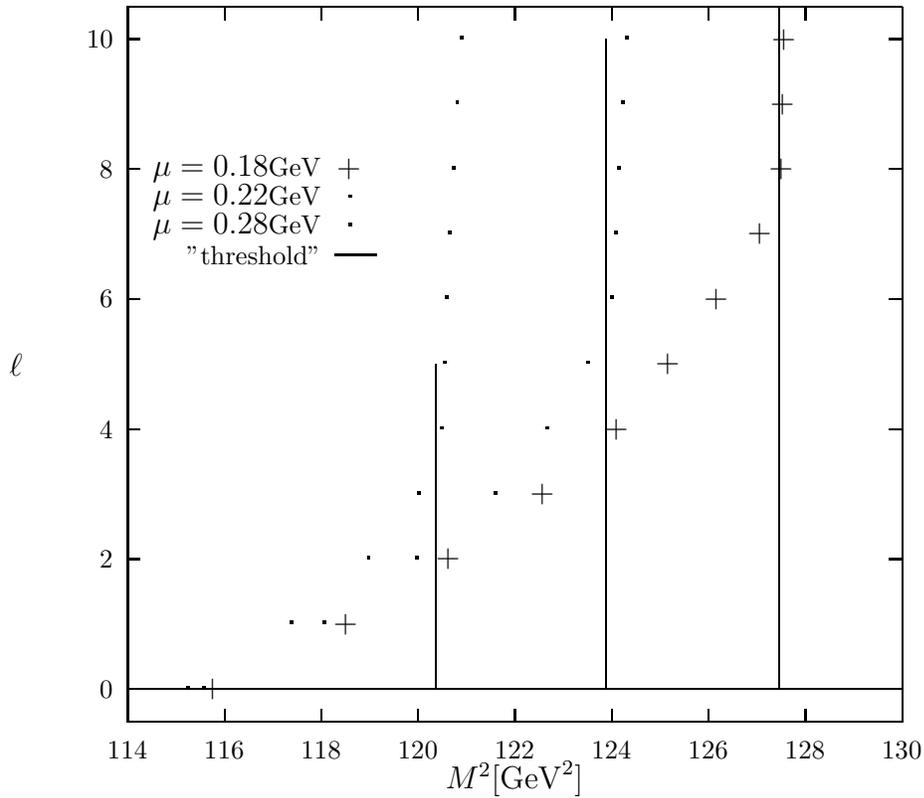

Our results are presented in Figs.~2-4. In Fig.~2 we show the parent
trajectory for the minimum, average and maximum screening $\mu$
extracted from the lattice study~\cite{lattice1}.  All three
trajectories clearly indicate a discontinuity in slope when the mass of
the bound state reaches the ionization level $(2M_b+\sigma / \mu)$.
We identify the states above the discontinuity as scattering states.  
Note that the maximum $\ell $  
rapidly decreases with increasing screening $\mu$. Daughter
trajectories exhibit the same behavior.
\begin{figure}[th!]
\input{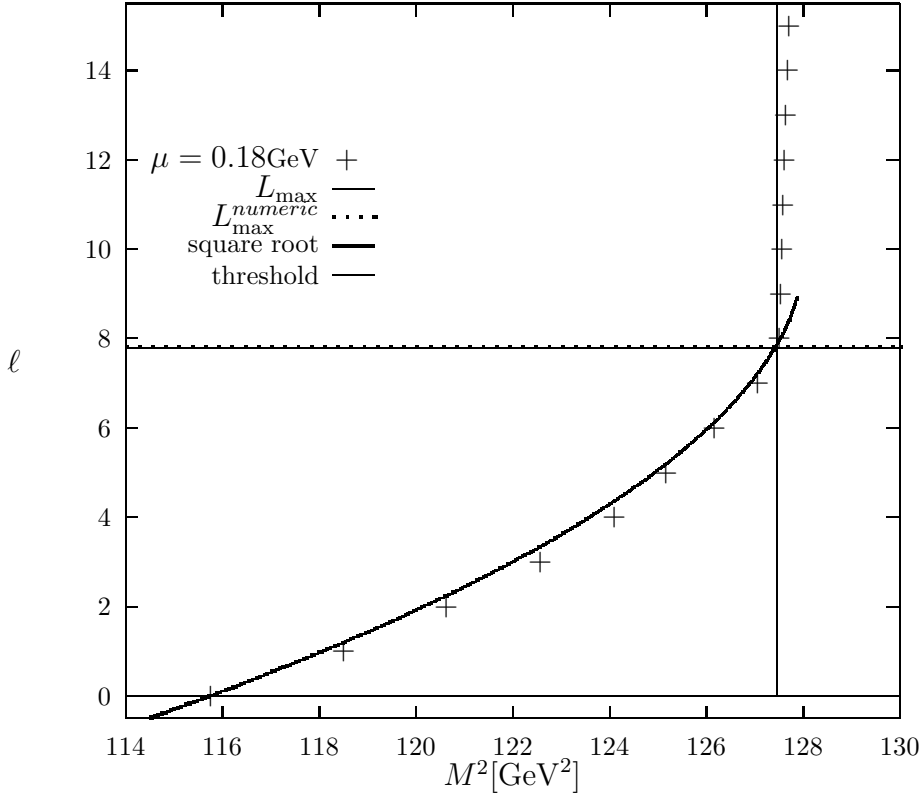}
\caption{The parent trajectory for bottomonium ($M_b=5.2$ GeV) with
the lattice unquenched potential $(2)$ for $\mu=0.18$ GeV, compared to a
``square-root'' trajectory with approximately the same slope at the
threshold. Horizontal lines represent $L_{max}$ according to Eq.
$(2)$, and according to a numerical extrapolation based on the data
points shown.}
\end{figure}

In view of the comments regarding the potential $(2)$, we argue that while we 
may
not have precisely identified the trajectory termination point, it
certainly occurs within a few states of the one below our model threshold. 
For
example, in Fig.~4,  the fourth daughter trajectory $\ell = 0$ state 
(corresponding
crudely to the $\Upsilon$(4S) state) may or may not be above
threshold.  Alternatively, there may be one or two states on succeeding
daughter trajectories (not shown) which should also be identified as
resonances with limited widths. In either case, the detail is not
significant to the point we wish to make here: That the calculated
trajectories (parent and daughter) have nonzero curvature, and that the
trajectories have a limiting value of $\ell$ (as well as $M^2$) beyond
which no reasonably definable states exist.

The maximum $\ell $ (see also Fig.~2 and Fig.~3) is consistent with
\begin{eqnarray}
L_{max} \simeq  {2\over{e}} \sqrt{\sigma M_b \over{\mu^3}} -1 -n \geq 0,
\end{eqnarray}
(where $n=0$ for the parent and $1\leq n \leq L_{max}(0)$ for the
daughter trajectories), which has been derived in~\cite{xy} for a
potential similar (but not identical) to $(2)$, viz.
\begin{eqnarray*}
V(r)=a+\frac{\sigma }{\mu }(1-e^{-\mu r}).
\end{eqnarray*}

Elsewhere, ``square-root'' Regge trajectories\footnote{
In practice, since for each hadronic Regge trajectories at most only three 
lowest lying states are known \cite{pdg}, the observed Regge trajectories 
can be well approximated by the well-known linear form. Nevertheless, some 
experiments indicate a nonzero curvature, for example, the nucleon trajectory 
\cite{nonlin1}, Pomeron trajectory \cite{nonlin2} and $a_2$ \cite{nonlin3}. 
Also, the straight line connecting $\rho$ and $\rho_3$ gives an intercept 
smaller than the actual physical value \cite{DL} (0.48 vs 0.55). Note in Figs.
2,3 that the trajectories are indeed approximately linear for $\ell \leq 3.$
This can be also seen analytically by the expansion of Eq. (4) for $t\ll T.$ }
have been studied as solutions to phenomenologically supported requirements 
of additivity of inverse slopes and intercepts \cite{BG} which are also 
consistent with the heavy quark limit \cite{future}. In view of
this success, it is interesting to see how much the QCD model trajectories
found here deviate from the phenomenologically plausible form. In
Fig.~4 we plot the parent trajectory for $\mu =0.18$ GeV, compared to a
``square-root'' trajectory that has the same slope at the termination
point,\footnote{This corresponds to  matching (approximately) the size
of the state, in addition to its mass and quantum numbers~\cite{future}. 
The parameters of the ``square-root'' trajectory matched here are:  
$\alpha(0)= -22.42$, $T=128.05$ GeV$^2$, and $\gamma =2.87$ GeV$^{-1}$.}
\begin{eqnarray}
\alpha (t)=\alpha (0)+\gamma \Big[ \sqrt{T}-\sqrt{T-t}\Big] .
\end{eqnarray}
where $\alpha (t)=\ell $ is the Regge trajectory, $\alpha(0)$ is the
intercept of the trajectory, $T$ is its termination point (note that we
allow $T$ to differ from the ionization threshold of the potential
$(2)$), and $\gamma$ is the universal asymptotic slope (i.e. $\alpha
(t)\sim \gamma (-t)^\nu ,\;|t|\rightarrow \infty $).  Since the masses
of the bound states with the lattice potential $(2)$ lie very close to
(i.e. in our case, within 1\% of) the ``square-root'' trajectory, all
of the phenomenology of ``square-root'' trajectories, such as
additivity of inverse slopes and intercepts, is applicable to these QCD
model bound state trajectories.

Note that  the parameters of the ``square-root'' form are free
parameters, a priori unknown, and must be fitted to the results of any
bound-state calculation. Therefore, even though the calculation
presented here should be expected to have uncertainties of up to 30\%,
due to the nonrelativistic approximation to the kinetic energy and due
to the use of  a static potential, it  remains highly plausible that
agreement of the exact QCD result with the phenomenologically supported
``square-root'' form can also be achieved (with similar parameter values).

As suggested by Eq.~$(3)$, $L_{max}$ should decrease by one unit for
each consecutive daughter trajectory.  Fig. 3 shows the parent and
daughter trajectories for $\mu=0.28$ GeV, and $L_{max}(n)$ given by
Eq.~$(3)$ for each of the trajectories. In this case, according to
$(3)$, there should be only three daughter trajectories, the last of
which consists of just one, $\ell=0$, state.

Our numerical results are in  agreement with Eq.~$(3)$. We
observe that $L_{max}$ decreases by roughly one unit; there are two
daughter trajectories with at least two bound states, and there is
indeed one more $\ell=0$ state near the threshold.

To summarize our findings: Using an unquenched lattice potential
$(2)$, we observe that there are a finite number of bound states
occupying Regge trajectories of a near-square-root form which terminate
in $\ell $, and a finite number of daughter trajectories. Quantitative
results (i.e. the value of $L_{max}$ and consequently, the number of
daughter trajectories) are sensitive to the exact value of the
screening parameter $\mu$, but the qualitative behavior is the same
over the entire range of $\mu$ allowed by~\cite{lattice1}.

We close these theoretical considerations with a few additional remarks. 

First, we would like to emphasize that the existence of the ionization
level (which is, in quark terms, an undesirable, but unavoidable feature of the
potential under consideration) does not guarantee termination of Regge
trajectories.  For example, in  QED, even though  there exists an
ionization level, there are infinitely many daughter trajectories, each
with, in principle, infinitely many bound states with integer $\ell $, in
contrast to finite $\ell $ in QCD. Since we do not know the solution to the
potential $(2)$ in an analytic form, we cannot conclude how many daughter 
trajectories it produces,  but it seems unlikely that the numerical 
methods would have entirely missed evidence of some singular behavior 
or accumulation of trajectories in a small region. Hence, it is likely 
that the number of daughter trajectories is finite also (see~\cite{xy}).

\begin{figure}[th!]
\input{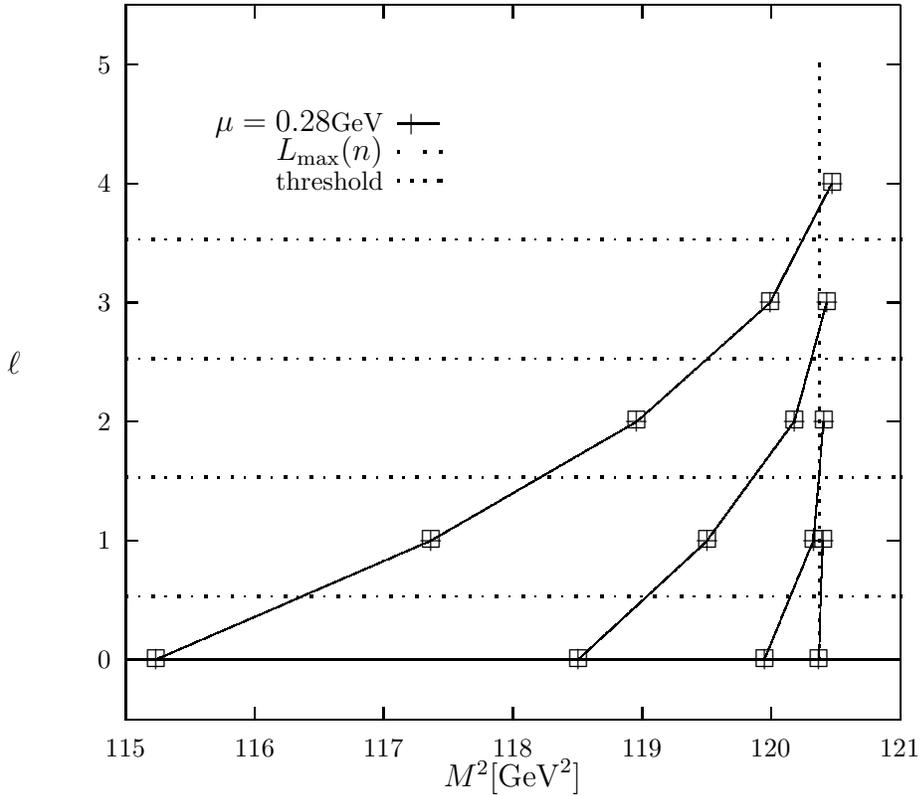}
\caption{The parent and daughter trajectories for the bottomonium
($M_b=5.2$ GeV) with the lattice unquenched potential $(2)$ for
$\mu=0.28$ GeV, together with $L_{\rm max}(n)$ according to Eq. $(3)$.
The line connecting the data points is to guide the eye.}
\end{figure}
Second, our results are supported  by recent lattice calculations which
observe flux tube breaking~\cite{lat1}, and the fact that there
have been many lattice simulations showing  flattening of the potential
due to pair production with much smaller errors than the original
lattice potential (used in our bound state calculation, see~\cite{lat2}
and references therein).

We therefore conclude that, in QCD, the flattening of the potential due
to pair creation produces termination of real part of  Regge trajectories. This 
is
in agreement with \cite{pet} where it is also suggested on different
grounds that Regge trajectories must terminate.

Finally, we address some of experimental consequences of this
effect, in particular, those regarding production processes,
spectroscopy and the proposed quark-gluon plasma.

Our finding that the bottomonium bound states lie within 1\% of the
``square-root'' trajectory may be useful for calculating cross sections
for production processes, such as photoproduction, where an explicit
form of the Regge trajectory exchanged is required. Moreover, evidence
from an analytical calculation for massless quarks, and model-dependent
studies~\cite{future}, strongly suggest that light quarkonia populate
near-square-root trajectories as well.

With regard to glueball spectroscopy: {\it If} one assumes that the
same mechanism which produces termination of Regge trajectories applies
to both gluonic bound states and quarkonia, {\it and} that they have
different thresholds (e.g. due to different color factors), then it is
conceivable that there is a range of masses where all $I=0$  states
are pure glueballs, or, perhaps, with small admixtures of the
lowest--lying heavy quarkonia. (A possible exception is the case of
mesonic molecules which, however, would be naively expected to have
much larger widths than ordinary resonances.) 

The finite number of bound states that we find also affects
understanding of the QCD phase transition from hadrons to the
quark-gluon plasma. In order to discuss the QCD phase transition, both
the hadron and the quark-gluon phases must be described by the
corresponding equations of state, each of which depends on the
corresponding degrees of freedom as functions of temperature. The
critical temperature is found as the point where the two curves
intersect. However, if the density of states in the hadron phase grows
without bound (which does occur for the case of linearly rising
trajectories \cite{Shuryak}), then at high enough temperatures the
hadron phase will be thermodynamically favored over the quark-gluon
plasma phase because the  effective number of degrees of freedom is
constant for the plasma.
This implies that the existence of the quark-gluon plasma would be
restricted to a limited range of temperatures (if it existed at all),
which is clearly unphysical. To be consistent with the concept of the
QCD phase transition, therefore, the effective number of degrees of
freedom in the hadron phase cannot grow indefinitely with temperature.
(Of course, QED also has an infinite number of bound states for hydrogen atom, 
which are truncated by medium effects. This kind of physics could also resolve 
the QCD problem. What we have presented here is an alternative solution.)

Obviously, our results that show the termination of the real part of 
trajectories at a 
certain energy threshold (in the calculation presented here 
$E_{th} \simeq 2M_b + {\sigma \over{\mu}}$ ), lead to a finite number of 
effective degrees of freedom\footnote{The scattering states, of course, do 
not contribute to this number.} that freeze out at $E = E_{th}.$ In this 
respect, the color screening discussed in this paper may be a manifestation 
of existence of the free phase of the hadron constituents -- quarks and 
gluons. The idea that color screening may in fact be the only mechanism 
responsible for deconfinement has been suggested in ref.~\cite{Satz}.

\vskip0.1in
\noindent
This research is supported by the Department of Energy under contract 
W-7405-ENG-36.

\end{document}